\documentstyle[prl,aps,multicol,epsfig]{revtex}
\begin{document} 
\draft 
\title{Excitation spectra of the two dimensional Kondo insulator}
\author{C. Gr\"ober and R. Eder}
\address{Institut f\"ur Theoretische Physik, Universit\"at W\"urzburg,
Am Hubland,  97074 W\"urzburg, Germany}
\date{\today}
\maketitle
\begin{abstract}
We present a Quantum Monte Carlo (QMC) study of the temperature dependent
dynamics of the two-dimensional (2D) Kondo insulator. Working at the so-called 
symmetrical point allows to perform minus-sign free QMC simulations. Study 
of the temperature dependence of the single-particle Green's function and 
the dynamical spin correlation function provides evidence for two 
characteristic temperatures, which we associate with the Kondo and coherence 
temperature, whereby the system shows a metal-insulator transition at the
coherence temperature. The data shows evidence for two distinct types
of spin excitations and we show that despite strong antiferromagnetic
ordering at low temperature the system cannot be described by 
spin-density-wave (SDW) theory.
\end{abstract} 
\begin{multicols}{2}
The periodic Anderson model (PAM) or its strong coupling version,
the Kondo lattice, may be viewed as the appropriate model
for describing intensively investigated classes 
of materials as the heavy electron metals \cite{Stewart,Fulde} and the
Kondo insulators \cite{Fisk,Schlesinger}. 
While the impurity case is well-understood\cite{Bickers,Keiter,Coleman,Kuramoto,Hartmann}
 and even amenable to exact
solutions\cite{Andrei,Wilson}, little is known about the lattice model.
Recently a considerable amount of
numerical results has been collected, albeit mainly
for the the one-dimensional (1D) case\cite{Moukuri,Tsutsui}. 
The main reason is that the frequently used
density matrix renormalization group (DMRG) method works best for 
1D systems. Only very recently finite-temperature exact diagonalization
results for the two-dimensional (2D) case\cite{Haule} became available.\\
The Quantum Monte-Carlo method\cite{Loh92,Hirsch85,Jarrel96} on the other hand,
can in principle treat systems of arbitrary dimension.
Here the limitation is in the notorious minus-sign problem,
which usually precludes the study of truly low temperatures.
However, by restricting oneself to the so-called symmetric
point the minus sign problem can be circumvented
and quite low temperatures be reached in numerical
simulations\cite{Vekic,Groeber}. In the present manuscript we want to present
data for the dynamics of the 2-dimensional (2D)
PAM with Hamiltonian
\begin{eqnarray}
H &=& -t\sum_{\langle i,j \rangle, \sigma} (c_{i,\sigma}^\dagger c_{j,\sigma}^{}  
+ H.c.) - V \sum_{i,\sigma} (c_{i,\sigma}^\dagger f_{i,\sigma}^{} + H.c.)
\nonumber \\
&-& \epsilon_{f} \sum_{i,\sigma} n_{i,\sigma} + U\sum_{i} f_{i,\uparrow}^\dagger 
f_{i,\uparrow}^{} f_{i,\downarrow}^\dagger f_{i,\downarrow}^{}.
\label{kondo1}
\end{eqnarray}
Here $c_{i,\sigma}^\dagger$  ($f_{i,\sigma}^\dagger$) creates a conduction 
electron ($f$-electron) in cell $i$, $n_{i,\sigma}$$=$$f_{i,\sigma}^\dagger 
f_{i,\sigma}^{}$ and $\langle i,j \rangle$ denotes nearest-neighbor lattice
sites. The so-called symmetric case corresponds to the special choice
$U=2\epsilon_f$. At `half-filling', i.e. electron density/unit cell
$n=2$, the Hamiltonian acquires particle-hole symmetry, whence there is no
more minus sign problem.\\
In a preceding work\cite{Groeber} we studied
the symmetric 1D case with interaction value $U=8.0 \, t$ and c-f hybridization 
$V=1.0 \, t$. For the lowest accessible 
temperature ($T=0.03 \, t$, less than $1\%$ of the conduction-electron bandwidth) 
the systems exhibits
insulating behavior with a gap in the single-particle spectrum. The c- and 
f-electrons seem to form a coherent all-electron fluid with composite c-f character 
of the low-lying one- and two-particle excitations, thus turning the system to
a `nominal' band insulator with a `Fermi-Surface' which covers
the entire Brillouin zone. Above the lower crossover temperature $T_{coh}$ (which we
identified with the so-called
coherence temperature in heavy Fermion systems) the c- and f-like features in the correlation functions 
are decoupled, indicating that the f-electrons `drop out' of the Fermi-surface.
The Fermi surface shrinks to one half of the Brillouin zone whence
the system becomes a metal and
the single-particle gap closes. At $T_{coh}$
the spin excitations of the f-electrons become localized
(visible as a dispersionless branch in the f-electron spin response) and 
the spin gap closes due to c-like spin excitations. Kondo 
resonance-like sidebands with f-character are seen in the single-particle
spectral function presumably as a sign of the formation of loosely bound singlets between
c- and f-electrons. At the higher crossover temperature $T_{K}$ these 
dispersionless f-like Kondo resonance-like sidebands in the single-particle 
spectrum disappear, whence we identify
it with the Kondo-temperature of the system. Above this temperature the 
single-particle spectral function of the c-electrons shows a very 
conventional tight-binding dispersion $-2t\cos(k)$, whereas the 
f-electrons show normal upper and lower Hubbard bands, i.e. the 
f-electrons do not participate in the low-energy physics.
While this is hard to establish numerically, it follows from
general theorems for 1D systems that there is no long range order
at any temperature.\\
This does not appear to be the case
in the 2D case studied previously with a smaller interaction $U=4.0 \, t$ by Vekic
{\it et al.} \cite{Vekic}.
There the ground state of the system is an insulator
with long-range AF order, a finite charge gap and gapless spin excitations 
for small values of $V^{2}/Ut$ (i.e. a Mott insulator). As $V^{2}/Ut$ 
increases, the long-range order is destroyed and spin-liquid behavior 
is found characterized by both a spin gap $\Delta_S$ and a charge 
gap $\Delta_C$ with 
$\Delta_{C} > \Delta_{S}$ (i.e. a Kondo insulator). Further increasing 
the hybridization $V$ then even leads to a band-insulating state with 
equal values of the spin and charge gap. The authors also found 
a Kondo resonance-like peak in the angle-integrated spectral density 
$D(\omega)$ for moderate temperatures and a gap for temperatures below.\\
Similar results concerning the ordered nature
of the ground state in 2D were obtained in a recent zero temperature QMC study of
the strong coupling version of the model by Assaad\cite{fakher}.
There, a quantum phase transition was found
to take place from an antiferromagnetically ordered phase
for $J/t < 1.5$, to a spin liquid phase for larger $J/t$.
Here $J$ denotes the Kondo exchange which 
would come out as $J= 8V^2/U$ in the strong coupling perturbation 
approach due to
Schrieffer and Wolf\cite{SchriefferWolf}. 
One would thus expect that in the Hamiltonian
(\ref{kondo1}) the transition occurs at $V^{2}/Ut\approx0.18$ 
in reasonable agreement 
with the results of  Vekic {\it et al.} for U=4.
\begin{figure}[htb!]
  \hspace{-0.25cm}\epsfig{file=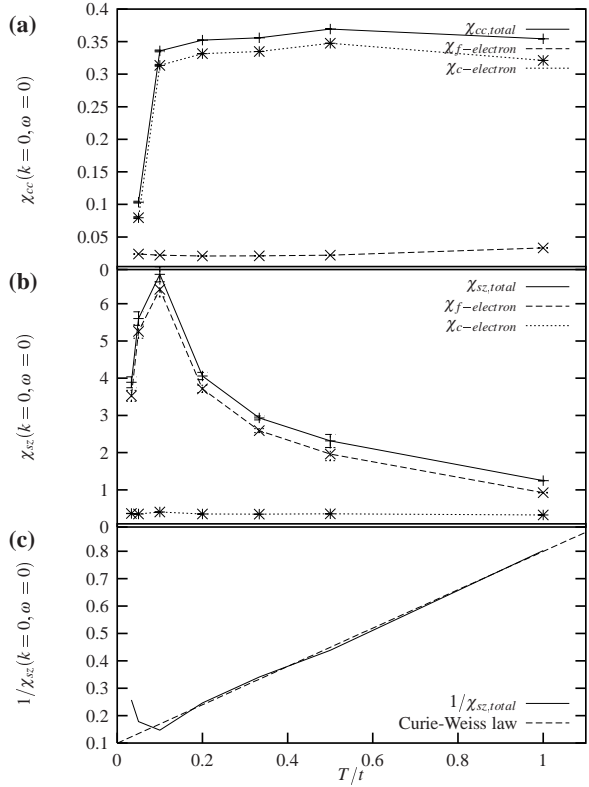, width=8.50cm}
  \narrowtext
  \caption[]{Susceptibilities of the 1D Kondo lattice with $U=8.0 \, t$ 
    and $V=1.0 \, t$ for different temperatures: (a) Charge susceptibility
    $\chi_{cc}$, (b) spin-susceptibility
    $\chi_{sz}$ and (c) inverse spin-susceptibility
    $1/\chi_{sz}$.}
  \label{fig1} 
\end{figure}
\noindent
In this work we study the 2D Kondo lattice with $6 \times 6$ unit cells 
using standard QMC techniques for the same interaction $U=8.0 \, t$ as in 
our previous work for the 1D case \cite{Groeber}. Again we restrict ourselves 
to the case of half filling, i.e. with two electrons/unit cell, again
at the symmetric point $\epsilon_{f}=U/2$ (i.e. $J/t$$=$$1$).
Due to the absence of the minus-sign problem we could reach temperatures as low as 
$T=0.05 \, t$, corresponding to $\approx 0.6\%$ of the conduction-electron 
(c-electron) bandwidth.\\
Based on the previous works\cite{Vekic,fakher} we expect that
due to our larger value for the interaction $U=8.0 \, t$ with a smaller ratio 
of $V^{2}/Ut$, an insulating ground state with 
long-range AF order to be stable. 
This is confirmed by our numerical results
for the spin and charge susceptibilities
$\chi_s^\alpha$ and $\chi_c^\alpha$ ($\alpha =c,f$ denotes the type of
electron probed). These are defined as
\begin{eqnarray}
\chi_{s}^\alpha &=&
\frac{\partial M}{\partial H}
= \int_0^{\beta T} \langle S_z^\alpha(\tau) S_z^\alpha(0) \rangle d\tau,
\nonumber \\
\chi_{c}^\alpha &=& \frac{\partial N}{\partial \mu}
= \int_0^{\beta T} \langle n^\alpha(\tau) n^\alpha(0) \rangle d\tau,
\end{eqnarray}
where $S_z^\alpha = \frac{1}{2}(n_\uparrow^\alpha- n_\downarrow^\alpha)$, 
$n^\alpha =(n_\uparrow^\alpha+ n_\downarrow^\alpha) $.
Figures \ref{fig1} and \ref{fig2}
show these susceptibilities for the 1D and 2D lattice. In both
cases the charge susceptibility for the conduction 
\begin{figure}[htb!]
  \hspace{-0.25cm}\epsfig{file=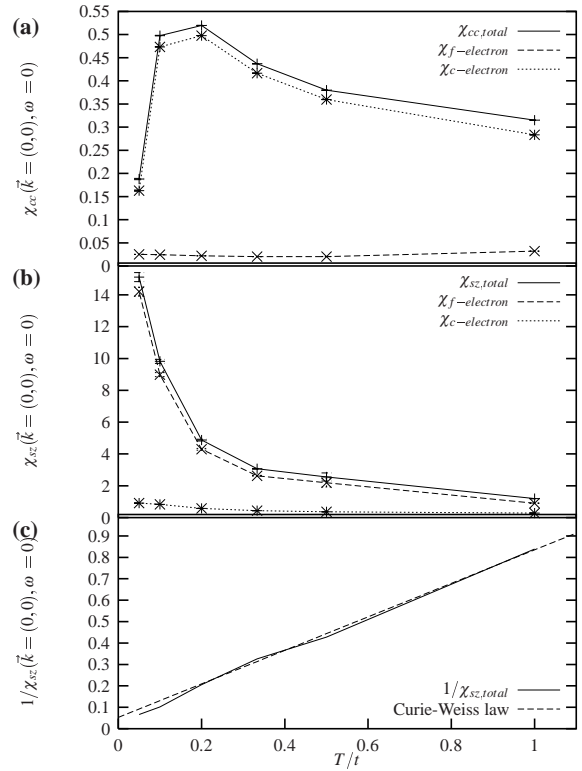, width=8.50cm}
  \narrowtext
  \caption[]{Susceptibilities of the 2D Kondo lattice with $U=8.0 \, t$ 
    and $V=1.0 \, t$ for different temperatures: (a) Charge susceptibility
    $\chi_{cc}$, (b) spin-susceptibility
    $\chi_{sz}$ and (c) inverse spin-susceptibility
    $1/\chi_{sz}$.}
  \label{fig2} 
\end{figure}
\noindent
electrons are considerably larger
than those for the $f$-electrons. This simply reflects the fact that the
$f$-electron subsystem is practically half-filled and the
strong Coulomb repulsion renders the $f$-electron system incompressible.
At high temperatures $\chi_c^c$ for the
1D system is more or less temperature independent, whereas it increases
with decreasing temperature for the 2D system.
This can be understood by assuming that in both cases
$f$ and $c$-electrons are decoupled at high temperature, as is
suggested by our previous study for the 1D model\cite{Groeber}.
The different temperature dependence then can be traced back to the
different free-electron density of states (DOS) in 1D and 2D.
Namely for free electrons with $\mu(T)=0$ one has
$\chi_c \approx  \frac{ \bar{\rho} }{k_BT}$
where $k_B$ is the Boltzmann constant and
$\bar{\rho}$ is the DOS averaged over a window of width
$k_B T$ around the chemical potential $\mu=0$.
This will be more or less constant for 1D, but strongly increasing with decreasing
$T$ in 2D, where one has a van-Hove singularity in the band center.
At temperatures below $0.10t$, however, $\chi_c$ drops sharply in both
1D and 2D. This is the behavior
expected for a band insulator and
indicates the opening of a gap in the fully interacting DOS,
because shifting $\mu$ does not change the particle number any more.
In both 1D and 2D we thus have evidence for the opening of a gap
in the single-particle spectrum.
\begin{figure}[htb!]
  \hspace{-0.25cm}\epsfig{file=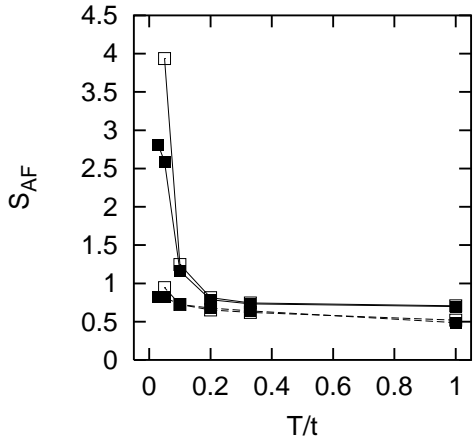, width=6.50cm}
  \narrowtext
  \caption[]{Antiferromagnetic structure factor $S_{AF}$ as a function
    of temperature for $f$-electrons (full line)
    and $c$-electrons (dashed line), calculated for
    the 2D $6\times 6$ cluster (open squares) and a 1D 16-cell
    chain (dark squares).}
  \label{fig3} 
\end{figure}
\noindent
Turning to the spin susceptibility we note that there the
dominant contribution comes from the $f$-electrons.
In both 1D and 2D $\chi_c^f$ can be fitted
roughly to a Curie-law at high temperatures. Deviations occur for smaller
$T\leq 0.20\, t$. It is also in this temperature range that
the difference between the 1D and 2D system becomes apparent:
the downturn of the spin-susceptibility $\chi_{sz}(T)$ at the 
coherence temperature $T_{Coh}$ for the 1D case signals a ground state with a spin-gap and
thus without long-range AF order, whereas the steady increase of the 2D spin-susceptibility 
$\chi_{sz}(T)$ suggests a ground state without a 
spin-gap and thus with long-range AF order in agreement with the data of Vekic {\it et al.} 
\cite{Vekic} and Assaad\cite{fakher}. We also note the overall agreement
of the susceptibilities
with the exact diagonalization data by Haule {\em et al.}\cite{Haule}.\\
A related quantity of interest is the static magnetic structure factor
\begin{equation}
S(\bbox{q}) = \langle S_{-\bbox{q}}^z   S_{\bbox{q}}^z \rangle.
\end{equation}
Figure \ref{fig3} shows the antiferromagnetic
structure factor $S_{AF}$, i.e. $S(\bbox{q})$ for $q=\pi$ in 1D and for $\bbox{q}=(\pi,\pi)$ in
2D. At high temperatures, $S_{AF}$ is $T$-independent and for the $f$-electrons
is close to the value of $0.75$ expected for a totally
uncorrelated spin-1/2 system. $S_{AF}$ for the $c$-electrons is
equally $T$-independent, but takes a smaller value presumably due to the reduction
of the $c$-moment by the charge fluctuations of a free-electron gas.
At low temperature $S_{AF}$ for the $f$-electrons in 2D increases
quite dramatically, indicating the tendency towards
long-range order.
The $c$-electron structure factor on the other hand
shows only a weak enhancement at low $T$: the ordering seems almost 
exclusively restricted to the $f$-electrons.
The  $S_{AF}$ in 1D also increases, but seems to saturate
at lower temperature.
In our preceding study in 1D\cite{Groeber} we found a spin correlation length of
$\zeta \approx 4.6$ at the lowest accessible temperature $T=0.03 \, t$ in
a ring of $16$ unit cells. 
The fact that $\zeta$ was already
quite significantly smaller than the cluster size indicated that
long range order does not develop in this case. A similar analysis
in 2D is difficult, because the clusters we can study have a much smaller
linear extent than in 1D, but it is visible that
the two systems behave quite different in their magnetic properties
at low temperatures. As we will see in a moment, this
difference at low temperatures has practically no bearing for
the temperature dependent
dynamics of the models, which are practically indistinguishable in 1D and 2D.
In particular we will see that the rather antiferromagnetic correlations
in the 2D system do not affect the single-particle spectra in any
noticeable way.\\
We start in Figure \ref{fig4} with a discussion 
of the temperature development of the angle-integrated spectral density of states 
$D(\omega)$ as a function of temperature
(see Ref. \cite{Groeber} for a definition of the various
correlation functions). At the highest temperature 
we studied, $T=1.00 \, t$, 
the f- and c-electrons are completely decoupled and
for both 1D and 2D system we observe nearly identical f-like upper and lower Hubbard bands at $\approx 
\pm U/2$, whereas the c-electrons show the density of states
expected for free electrons in the respective dimension.
This decoupling of the c- and the f-electron physics is also 
visible in the momentum resolved single-particle spectral function $A(\bbox{k},\omega)$ plotted 
in Figure \ref{fig5} 
(now and for the rest of this work only for the 2D case; the similarity
between 1D and 2D is nevertheless striking and we refer the readers to the plots in our 
preceding publication \cite{Groeber}). At temperature $T=1.00 \, t$ the c-electrons show a very 
conventional free tight-binding dispersion, whereas the f-electrons obviously do not 
participate in the low-energy physics at all. The momentum resolved 
spin-correlation function $S(\bbox{k},\omega)$ (see Figure \ref{fig6}) shows a 
free-electron-like particle-hole continuum in the c-electron case and a practically 
dispersionless branch in the f-like spectrum, identifying the magnetic f excitations as 
practically immobile.
\begin{figure}[htb!]
  \hspace{-0.25cm}\epsfig{file=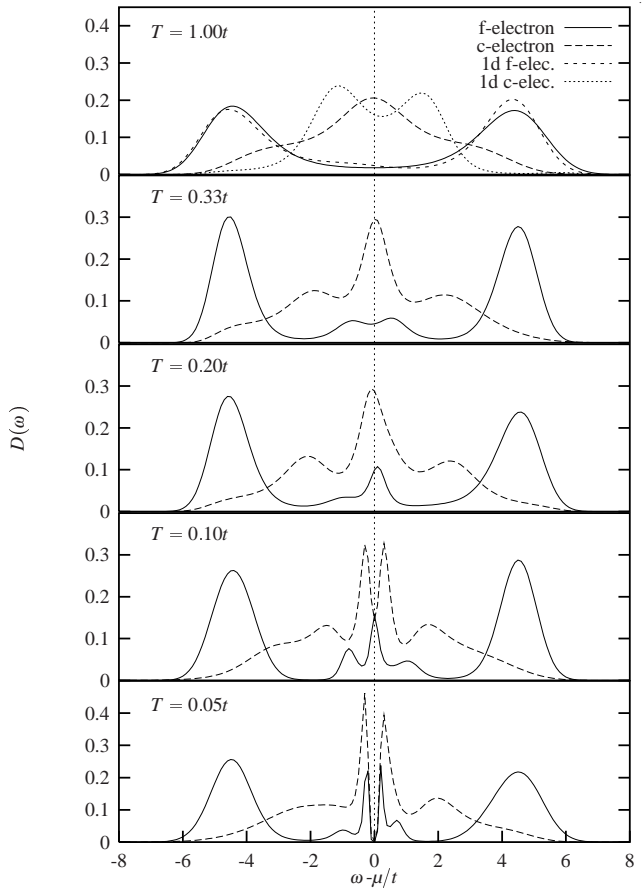, width=8.50cm}
  \narrowtext
  \caption[]{Angle-integrated spectral density $D(\omega)$ of the 2D Kondo
    lattice with $U=8.0 \, t$ and $V=1.0 \, t$ for different temperatures. 
    The 1D Kondo lattice density of states $D(\omega)$ is shown at temperature 
    $T=1.00t$ for comparison.}
  \label{fig4} 
\end{figure}
\noindent
Lowering the temperature to $T=0.33 \, t$ the density of states $D(\omega)$ 
shows f-like spectral weight around $\mu$ which forms a single peak 
right at $\mu$ as the temperature is lowered further to $T=0.20\, t$. We interpret this as a sign of the 
formation of loosely bound singlets between f- and c-electrons, i.e.
the $f$-electrons  now start to participate in the low-energy physics. At this temperature
the c-electrons also start to deviate from the free tight-binding dispersion with a broadened
and slightly split peak right at $\mu$ for momentum $\bbox{k}=(\pi,0)$ and with replicas of
the tiny f-like ``foot'', best visible at the neighboring momenta $\bbox{k}=(3\pi/4,0)$ and 
$\bbox{k}=(\pi,\pi/4)$, but also for $\bbox{k}=(\pi/2,\pi/2)$. At $T=0.20t$ the 
dynamical spin-correlation function $S(\bbox{k},\omega)$ of the f-electrons
shows still a dispersionless branch, i.e. the magnetic f excitations are still
immobile, but now with a much more narrow width.
In addition broad c-like 
low-energy `humps' are being
split off from the c-like electron-hole continuum, 
best visible along $(0,0)\rightarrow (\pi,\pi)$.\\
Proceeding to the temperature $T=0.10t$ the f-like density of states in Figure
\ref{fig4} shows the formation of side bands at $\approx \pm 1 \, t$ with the c-like peak right at
$\mu$ starting to split as precursor of the formation of the single-particle gap. 
\begin{figure}[htb!]
  \hspace{-0.25cm}\epsfig{file=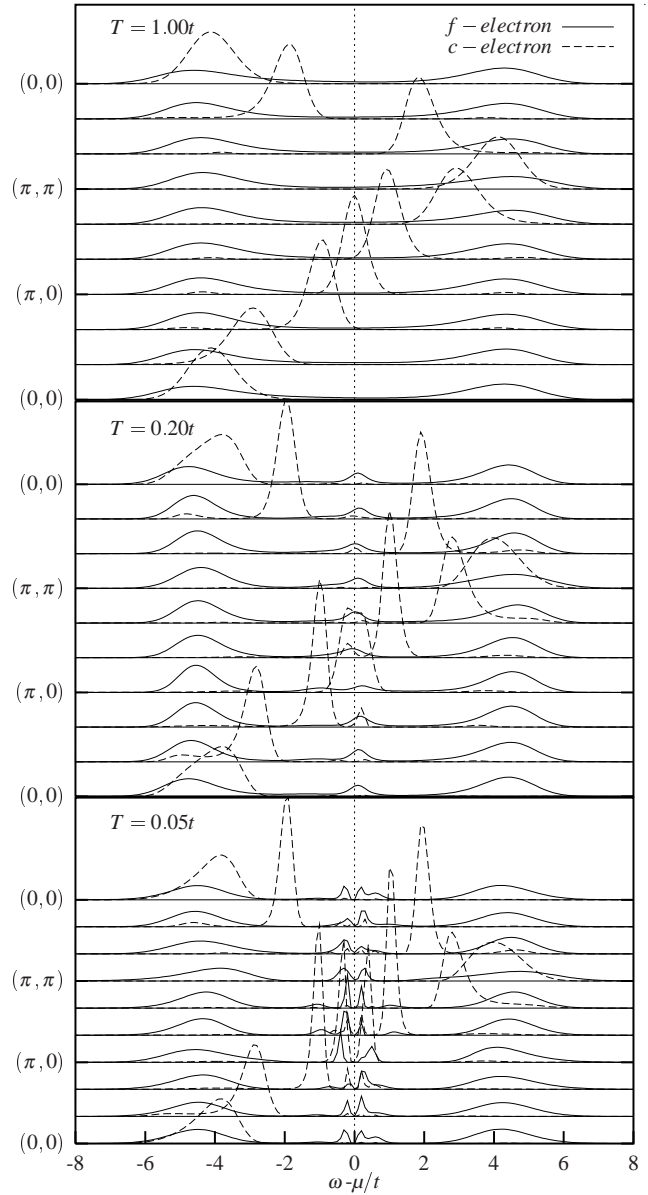, width=8.50cm}
  \narrowtext
  \caption[]{Momentum resolved single-particle spectral function 
    $A(\bbox{k},\omega)$ with same parameters as in Figure \ref{fig4}
    for different temperatures.}
  \label{fig5} 
\end{figure}
\noindent
At the lowest 
temperature $T=0.05 \, t$ the c-electron density at higher energies 
is still consistent with a standard 2D 
tight-binding density of states (the maxima
at $\approx \pm 2 \, t$ are a finite-size effect - we have checked that they
appear also when a free tight binding band is simulated
in a finite cluster). At low energies, however, $D(\omega)$
shows a clear gap around $\mu$. This demonstrates the insulating nature of the 
ground state, in agreement with the
behavior of the charge susceptibility $\chi_{cc}$. 
The f-electrons show the high-intensity upper and lower Hubbard bands at 
$\approx \pm U/2$, but now also very sharp low-energy peaks at the gap edges and in addition
the two side bands at $\approx \pm 1 \, t$. 
These sidebands are best visible in the spectral function $A(\bbox{k},\omega)$ of the f-electrons 
shown in Figure \ref{fig5} and are accompanied by a further change in the spectral function of 
the c-electrons: The $\cos(k_{x})+\cos(k_{y})$ dispersion of the
c-electrons now shows a gap at $\bbox{k}^{0}_{F}=(\pi,0)$.
\begin{figure}[htb!]
  \hspace{-0.25cm}\epsfig{file=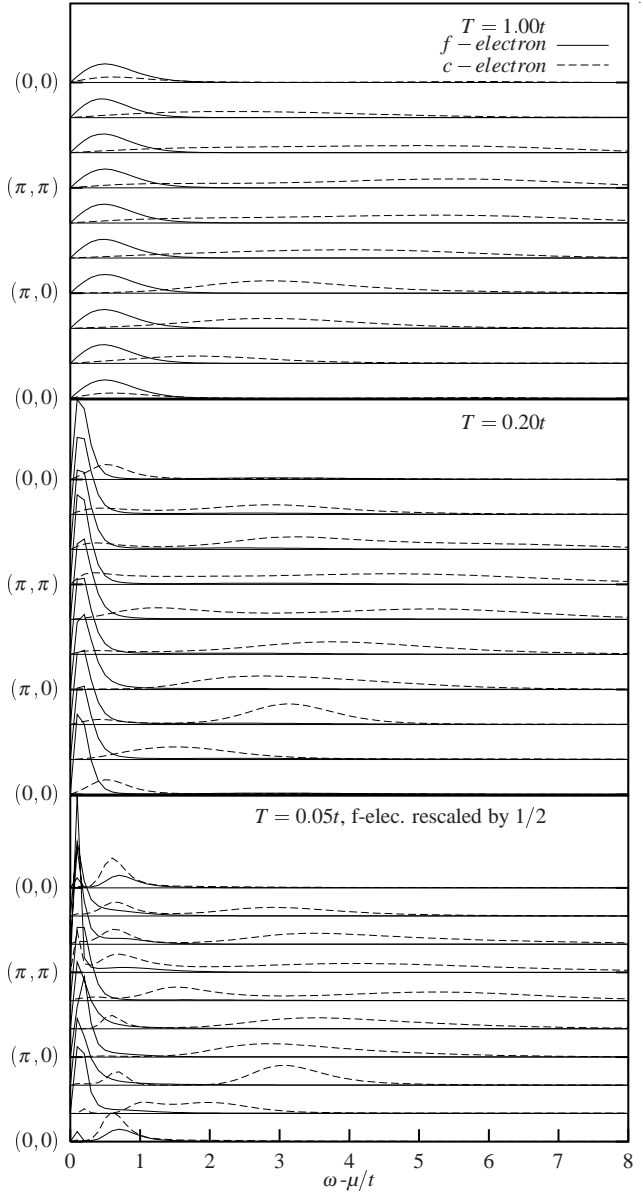, width=8.50cm}
  \narrowtext
  \caption[]{Momentum resolved dynamical spin-correlation function 
    $S(\bbox{k},\omega)$ with same parameters as in Figure \ref{fig4}
    for different temperatures.}
  \label{fig6} 
\end{figure}
\noindent
In the $6\times 6$ cluster this is the only momentum on the
Fermi surface for noninteracting conduction electrons, but one may expect
that in a larger system the gap is uniform in $\bbox{k}$-space.
The states with the lowest excitation energies
now are practically dispersionless bands with small, $f$-like weight.
Obviously the $f$-electrons now fully participate in the
single-electron states close to $\mu$. Assuming the validity of the Luttinger theorem
with the number of electrons being given by $c$- {\em and} $f$-electrons together
would also give an obvious explanation for the insulating nature of the ground state -
at electron density $n=2$ the system should be a band insulator.
With the exception of the $f$-like side bands the single particle spectrum
agrees very well with theoretical 
predictions based on an `expansion around an singlet vacuum'\cite{oana}.\\
The dynamical spin-correlation 
function $S(\bbox{k},\omega)$ of the f-electrons at $T=0.05\ t$ shows an
intense branch of low-energy excitations with a weak dispersion.
Its spectral weight has a sharp maximum at $\bbox{k}=(\pi,\pi)$, 
consistent with the strong AF correlations
for the $f$-electrons. In the c-electrons' spin correlation function
there appears a practically dispersionless low energy excitation
at $\omega \approx 0.50t$. In some cases one can see that the
$f$-electron spectrum also shows a weak hump at the position
of this dispersionless excitation, indicating that this excitation
has a mixed $f$-$c$ character. The sharp $f$-like low energy mode also has
some admixture of $c$-weight at $(\pi,\pi)$ - the spin excitations at
low energies thus have both $f$ and $c$ character, which shows again
that the two types of electron have merged to form a common
all-electron fluid. \\
As a  surprising result
the temperature development of the dynamical correlation functions
resembles in considerable detail that seen previously in 1D\cite{Groeber} -
even the `crossover-temperatures' in the spectral function
are practically the same in 1D and 2D. In fact, with the sole exception
of the different form of the c-electron DOS at higher energies, the angle integrated
spectra in Figure \ref{fig4} are practically indistinguishable from their
1D counterparts in Figure 1 of Ref. \cite{Groeber}, and this holds true for
each individual temperature. Assuming that this
is not coincidence for the special set of parameters
we are using, the characteristic temperatures for
the 1D and 2D models with identical parameters thus are at least very close to
one another. This would be hard to understand if we
were to assume that the characteristic
temperatures of the model depend sensibly e.g. on the c-electron density
of states at the Fermi energy, $\rho_F$ - this would be
singular (or at least significantly larger on a discrete lattice) 
for the 2D case.
In the case of the Kondo-impurity, where the
Kondo temperature is $T_{K} \propto \exp(-\frac{1}{\rho_F J})$, we would thus expect
a very different temperature evolution.
To be more quantitative, we estimated the impurity model Kondo temperatures
for {\em finite systems}. To better take into account the
particle-hole symmetry we assume that the
transformation to the strong coupling model has been
performed, i.e. we use
\begin{equation}
H = -t\sum_{\langle i,j \rangle, \sigma} (c_{i,\sigma}^\dagger c_{j,\sigma}^{}  
+ H.c.) + J \sum_{\bbox{k},\bbox{k}'}
\vec{S} \cdot c_{\bbox{k},\sigma}^\dagger  \vec{\tau}_{\sigma,\sigma'}
c_{\bbox{k}',\sigma'},
\end{equation}
where $\vec{\tau}$ denotes the vector of Pauli matrices,
and $\vec{S}$ the spin operator for the $f$-like impurity.
The value of $J=8V^2/U$ thereby takes into account the
fact that the exchange processes can occur both via
an empty or a doubly occupied $f$-level.
We then  use the variational trial state by
Yoshimori\cite{yoshimori}, which yields a self-consistency equation
for the binding energy $E_S$ of the Kondo singlet:
\begin{equation}
1 - \frac{3J}{4N} \sum_{\bbox{k} \in unocc.}
\frac{1}{\epsilon_{\bbox{k}} - E_S}=0.
\end{equation}
To solve the equation, we take the true discrete $\bbox{k}$-meshes
of the $16$-site chain and the $6\times 6$ lattice, which
presumably takes finite-size effects into account to some degree.
We then find for our parameters the singlet binding energies
$E_S=-0.0624\;t$ in 1D and $-0.1219\;t$ in 2D.
The Kondo temperatures in the impurity model thus would
differ by roughly a factor of $2$, but in the lattice model
the characteristic temperatures are more or less indistinguishable.
Assuming that the (near)-identity between the characteristic
temperatures in 1D and 2D is not just a coincidence for the
specific parameter set we are studying, this suggests that the
characteristic temperatures for the lattice system have little or no
relationship with those for the impurity model.
\begin{figure}[htb!]
\vspace{-0.5cm}
  \hspace{0.5cm}\epsfig{file=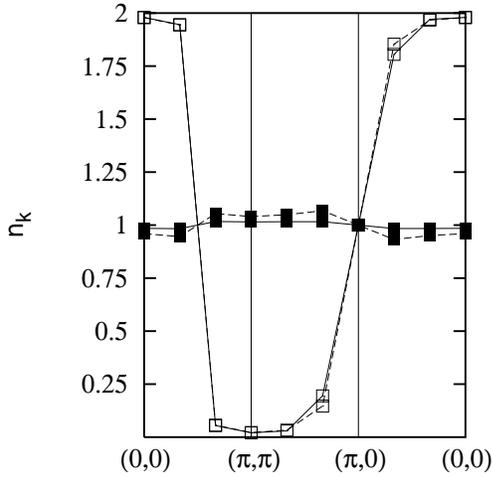, width=6.50cm}
\vspace{0.5cm}
  \narrowtext
  \caption[]{Momentum distribution for $f$-electrons (full squares)
    and $c$-electrons (light squares) at $T=0.33t$ (full lines) and at
    $T=0.05t$ (dashed lines) along high symmetry lines of the Brillouin zone
    for the $6\times 6$ cluster.}
  \label{fig7} 
\end{figure}
\noindent
A further surprising feature of the results is the absence of any 
sign of antiferromagnetism in the single-particle spectra: 
the $\bbox{k}$-resolved spectra for the $c$-electrons
do not show any indication of antiferromagnetic umklap bands
(for the $f$-electrons it is impossible to
make such a statement because the $f$-like bands are all more or less
flat). Figure \ref{fig7} shows the momentum distribution
$n^\alpha_{\bbox{k}} = \sum_\sigma \langle \alpha_{\bbox{k},\sigma}^\dagger 
\alpha_{\bbox{k},\sigma} \rangle$ 
($\alpha=c,f$) at high temperature $T=0.33t$, where no
noticeable enhancement of $S_{AF}$ could be seen in Figure  \ref{fig3} 
and at the
lowest temperature $T=0.05t$ where $S_{AF}$ shows strong signatures
of antiferromagnetism.
The most prominent feature is the `Pseudo Fermi surface' for the
$c$-electrons, i.e. a sharp drop of $n^c_{\bbox{k}}$ which occurs
at the Fermi surface of the {\em unhybridized} conduction electrons.
This is familiar from numerical studies of the 1D model\cite{Moukuri,Tsutsui}
and can be explained theoretically by `expansion around the singlet vacuum'\cite{oana}.
Surprisingly, $n^c_{\bbox{k}}$ is almost indistinguishable,
the only change at low temperature being a `sharpening'
of the distribution near $(\pi,0)$.
Similarly, $n^f_{\bbox{k}}$ develops some structure at low
temperature and is completely flat at high temperature
(the latter again shows the decoupling of $c$ and $f$-electron at high
temperatures). We note that both
changes are in fact opposite to what one would expect
on the basis of a conventional spin-density-wave (SDW)
picture: there, the static SDW
would provide an additional potential $V(\bbox{R}_i)= V_{AF} e^{i \bbox{Q}\cdot \bbox{r}}$
which would tend to mix the single particle states
$\alpha_{\bbox{k},\sigma}^\dagger$ and $\alpha_{\bbox{k}+\bbox{Q},\sigma}^\dagger$.
Any difference between $n^\alpha_{\bbox{k}}$ in the
inner part and the outer part of the antiferromagnetic Brillouin zone
should thus be {\em reduced} by the antiferromagnetic ordering.
By contrast, the actual data show that the structures in $n^\alpha_{\bbox{k}}$
actually sharpen up in the ordered state. \\
To be more quantitative, let us briefly discuss inhowmuch 
an SDW-like mean-field theory might explain our results.
We approximate the interaction term for the $f$-electrons:
\begin{eqnarray}
U n_{i,\uparrow} n_{i,\downarrow}
&\rightarrow&
U \langle n_{i,\uparrow} \rangle  n_{i,\downarrow}
U \langle n_{i,\downarrow} \rangle  n_{i,\uparrow}
\nonumber \\
&=& \frac{n_f U}{2} + \sum_\sigma
\frac{\sigma U m}{2} e^{i \bbox{Q} \cdot \bbox{R}} n_{i,\sigma}.
\end{eqnarray}
Here $n_f$ is the average density of $f$-electrons/site, and
particle-hole symmetry at half-filling implies $n_f$$=$$1$.
The parameter $m$ is the staggered magnetization of $f$-electrons.
Introducing the vector 
$C = (c_{\bbox{k},\sigma}, c_{\bbox{k}+\bbox{Q},\sigma}
f_{\bbox{k},\sigma}, f_{\bbox{k}+\bbox{Q},\sigma})$, the Hamiltonian
then takes the form
\begin{equation}
H = \sum_{\bbox{k} \in AFBZ} \sum_\sigma
C_{\bbox{k},\sigma}^\dagger H_{\bbox{k},\sigma} C_{\bbox{k},\sigma}^{}
\end{equation}
with the matrix
\begin{equation}
 H_{\bbox{k},\sigma} =
\left(
\begin{array}{c c c c}
\epsilon_{\bbox{k}}, & 0, &-V,& 0\\
0,& -\epsilon_{\bbox{k}} &, 0,&-V\\
-V, & 0,& 0,& \frac{-\sigma m U}{2}\\
0, & -V,& \frac{-\sigma m U}{2},& 0\\
\end{array} \right).
\label{matrix}
\end{equation}
For our parameter values, self-consistent
calculation of the staggered magnetization $m$
yields the value $m \approx 0.95$.
The band structure obtained in this way is shown in Figure
\ref{fig8}. It can be understood by considering the limit
$V/U \ll 1$ where we expect first of all
dispersionless $f$-like bands at $\pm U/2$ (obtained by
diagonalizing the $2\times2$ matrix in the lower right corner of
(\ref{matrix})). Eliminating these bands from
the Hamiltonian by canonical perturbation theory generates a coupling matrix
element between
$c_{\bbox{k},\sigma}$ and $c_{\bbox{k}+\bbox{Q},\sigma}$
which is equal to $-2V^2/U$.
We thus expect for the $c$-electrons a band structure
which resembles the SDW-approximation for the
single-band Hubbard model, with the relatively small SDW gap parameter
$2 V^2/U$.
The dispersionless $f$-like bands close to $\mu$,
which can be seen rather clearly in the numerical spectra
in Figure \ref{fig4},
are not at all reproduced by this approach, whence
a simple SDW-like mean-field theory is clearly
inadequate to explain the band structure generated by
the QMC simulation.
\begin{figure}[htb!]
  \hspace{0.5cm}\epsfig{file=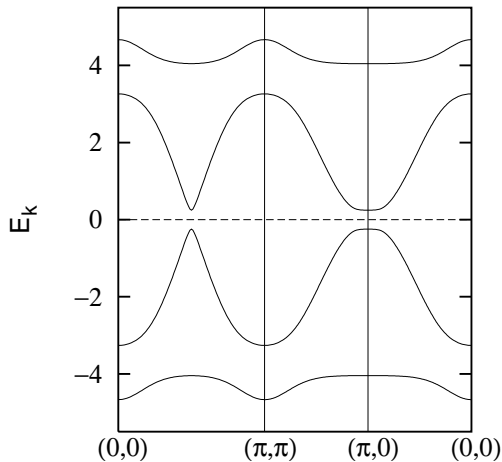, width=6.50cm}
\vspace{0.5cm}
  \narrowtext
  \caption[]{Self-consistent SDW band structure for the same
    parameter values as Figure \ref{fig5}.}
  \label{fig8} 
\end{figure}
\noindent
The above results allow some conclusions
concerning the spin excitations responsible for the
ordering if we consider the possible states of a single cell.
If we restrict ourselves to low-energy states, 
the most likely states are those with precisely
one $f$-electron/unit cell. These would be
\begin{eqnarray}
|i,1,\sigma\rangle &=& f_{i,\sigma}^\dagger |0\rangle,
\nonumber \\
|i,3,\sigma\rangle &=& c_{i,\uparrow}^\dagger
c_{i,\downarrow}^\dagger f_{i,\sigma}^\dagger |0\rangle,
\nonumber \\
|i,2,0\rangle &=&
\frac{1}{\sqrt{2}} (
c_{i,\uparrow}^\dagger
f_{i,\downarrow}^\dagger -
c_{i,\downarrow}^\dagger f_{i,\uparrow}^\dagger ) |0\rangle,
\nonumber \\
|i,2,1\rangle &=&
\frac{1}{\sqrt{2}} (
c_{i,\uparrow}^\dagger
f_{i,\downarrow}^\dagger +
c_{i,\downarrow}^\dagger f_{i,\uparrow}^\dagger ) |0\rangle.
\end{eqnarray}
These states have a spin of 1/2 ($|i,1,\sigma\rangle$ and
$|i,3,\sigma\rangle$), 0 ($|i,2,0\rangle$) or 1 ($|i,2,1\rangle$).
Above, we have defined the $S_z=0$ component of the triplet, but
there are also states with $S_z=\pm1$.\\
Then, one can envisage two very different
types of spin excitations:\\
a) One could flip the spin of one of the charged cells,
e.g. convert $|1,\sigma\rangle \rightarrow |3,\bar{\sigma}\rangle$.
In these states the spin of the cell is carried
exclusively by the $f$-electron. This means that
this type of spin excitation can be created only
by acting with the $f$-spin operator. The $c$-electrons'
spin operator obviously cannot `touch' these types of single-cell
states.\\
b) Alternatively, one could convert the singlet into one of the three
components of the triplet. In this case, the total spin of the
cell is carried by both, the $c$ and the $f$-electron. This 
type of excitation therefore can be generated by
both, the $f$- and the $c$-spin operator.\\
Since the two types of spin excitations have identical
quantum numbers, one might expect that the
`true' spin excitations of the system are a mixture of the
two. However, our data suggest that the mixing between the
two types of excitation is indeed quite weak: for example the
fact that it is predominantly the $f$-like structure factor
$S_{AF}$ which shows ordering in Figure \ref{fig3}
indicates that the AF ordering is predominantly due to ordering
of the $f$-spins in singly and three-fold occupied cells.
Let us for example define
\begin{equation}
|i,\sigma\rangle =
\cos(\Theta) |i,2,0\rangle + 
\frac{\sin(\Theta)}{\sqrt{2}} ( |i,1,\sigma \rangle +
|i,3,\sigma \rangle ),
\end{equation}
and (introducing the two sublattices $A$ and $B$)
\begin{equation}
|\Psi \rangle = {\cal P}_{2N} \prod_{i \in A} |i,\uparrow \rangle
\prod_{i \in B} |i,\downarrow \rangle,
\end{equation}
where ${\cal P}_{2N}$ projects onto states with
precisely $2N$ electrons and $z$-spin $S_z=0$. The state $|\Psi \rangle$
then has N\'eel order with an ordered moment
$\propto \sin^2(\Theta)$ in the $f$-system, but
no order whatsoever in the $c$-electron system.
We do not claim that the state $|\Psi \rangle$ has much to do with the
ground state of the lattice model - it demonstrates, however, that
by using singly and three-fold occupied cells it is
indeed possibly to construct states where only the $f$-electrons
do order. This suggests that the ordering
is driven by the charged single-cell states
$ |i,1,\sigma \rangle$ and $|i,3,\sigma \rangle$ (which should be
modeled as effective Fermions\cite{oana}) and not the singlet-triplet
excitation. This assumption also explains immediately ,why the
$c$-electrons momentum distribution (see Figure \ref{fig7})
and spectral density (see Figure \ref{fig5})
is so remarkably unaffected by the AF order at low
temperature.\\
Rather clear evidence for these two types of spin excitations
is also provided by the dynamical spin correlation function
at low temperature. There, we have seen at the lowest
excitation energies 
an almost purely $f$-like spin excitation with some relatively
small admixture of $c$-weight near $(\pi,\pi)$. This would suggest
that this mode has the character of a particle-hole excitation
carried by the charged spin-1/2 cells, 
$|i,1,\sigma\rangle$ and $|i,3,\sigma\rangle$.
At a somewhat higher excitation energy, the data
showed a practically dispersionless mode
with strong $c$-character but also some $f$-character - this 
might correspond to the singlet-triplet excitation.
The weak admixture of $c$-weight at $(\pi,\pi)$ in the low energy
spin excitation then is a measure for the mixing between the two
types of spin excitation.\\
In summary, we have presented a QMC study of the temperature dependent dynamics
in the 2D Kondo lattice model at half-filling. Working at the
symmetric point, $U=2\epsilon_f$, we could avoid the minus-sign problem
and study the evolution with temperature down to extremely low temperatures.
As was the case in 1D\cite{Groeber}, we could identify two characteristic
temperatures. The lower one of these, which we identify with the
experimental coherence temperature, is associated with the metal-insulator transition in the half-filled case, $n=2$.
Below $T_{coh}$ the $f$-electrons merge with the conduction electrons
to form a coherent all-electron fluid. The $f$-electrons consequently
participate in the Fermi surface volume, whence the system would be
a `nominal' band insulator even in the absence of any antiferromagnetic
order. In 2D the ground state is known\cite{Vekic,fakher}
to have antiferromagnetic order, which in principle could
turn even the {\em unhybridized} conduction electron system
into an insulator, due to the reduction of
the Brillouin zone by a factor of two.
However, our data do not show any indication of antiferromagnetic
ordering in the c-like single-particle spectrum
or the c-like momentum distribution, so that
the `effective' SDW-potential felt by the c-electrons due to the
ordering of the $f$-electrons must be extremely
weak. Moreover, the band structure obtained
from simple SDW mean field calculation is qualitatively different from
the numerical results, and in particular would lack the
dispersionless low energy band with $f$-character, which are
typical of the band structures in both 1D and 2D.
By analogy with the 1D system we thus conclude that
antiferromagnetism is not essential for the
insulating nature of the ground state (as can be seen also from the
fact that the strong coupling model remains an insulator
in the spin-liquid phase\cite{fakher}), but that it is the
`merging' of the $c$- and $f$-electron systems which drives the
metal-insulator transition with decreasing temperature.\\
Increasing the temperature beyond $T_{coh}$ the $f$-electrons
drop out of the Fermi surface volume, which results in the metal-insulator
transition. The $f$-electrons do no longer participate in the
low energy physics, the first ionization states are
purely $c$-like and the $f$-spin excitation becomes localized.
In the single-particle spectrum we can identify dispersionless
$f$-like sidebands, well separated from the chemical potential.
This suggests that local singlets between $f$ and $c$-electrons still
exist, but no phase coherence between the singlets is established.
Finally, at the highest temperatures the $f$-electrons disappear completely
from the low energy single-particle spectrum, and all that remains
are two $f$-like Hubbard bands and the standard free-electron band for the
$c$-electrons. We associate the temperature where this complete decoupling
occurs as the analogue of the Kondo temperature.\\
A surprising feature of the results is the close similarity
with those obtained for the 1D system\cite{Groeber}. In particular, the characteristic
temperatures are more or less identical (to the accuracy to which these
`crossover' temperatures can be assigned) in 1D and 2D.
Together with the very different nature
of the single-particle DOS around the band center in the
1D and 2D tight-binding bands
this seems to indicate that in the lattice case there is no significant
dependence of the characteristic temperatures on the density of states
at the Fermi energy. \\
At low temperatures our data in 2D are consistent with
long range antiferromagnetic order, in agreement with 
previous work by Vekic {\em et al.} and by Assaad\cite{fakher}. 
However, our data
do not show any indication of `antiferromagnetic symmetry'
in the single particle spectra, in particular there are
no distinguishable antiferromagnetic umklap bands in the
spectra. This again indicates that an SDW-like treatment of the
antiferromagnetic state is not adequate.\\
This work was supported by DFN Contract No. TK 598-VA/D03, by BMBF (05SB8WWA1), 
and by computing resources from HLRS Stuttgart and HLRZ J\"ulich.

\end{multicols}
\end{document}